\documentclass[sn-nature]{sn-jnl}% Style for submissions to Nature Portfolio journals
\usepackage{graphicx}%
\usepackage{multirow}%
\usepackage{multicol}
\usepackage{amsmath,amssymb,amsfonts}%
\usepackage{amsthm}%
\usepackage{mathrsfs}%
\usepackage[title]{appendix}%
\usepackage{xcolor}%
\usepackage{textcomp}%
\usepackage{manyfoot}%
\usepackage{booktabs}%
\usepackage{algorithm}%
\usepackage{algorithmicx}%
\usepackage{algpseudocode}%
\usepackage{listings}%

\usepackage{gensymb}
\usepackage{geometry}
\usepackage{gensymb}
\usepackage{graphicx}
\usepackage{subcaption}
\usepackage{amsmath}
\usepackage{lineno}
\usepackage{enumerate}
\usepackage{wrapfig}
\usepackage{multirow}
\usepackage{soul}
\usepackage{color}
\usepackage{parskip}

\begin{document}

\title[Article Title]{Froude number scaling unifies impact trajectories into granular media across gravitational conditions}            


\author[1]{\fnm{Peter M.} \sur{Miklav\v ci\v c}}
\email{pmiklavc@ur.rochester.edu}
\equalcont{These authors contributed equally to this work.}

\author[1]{\fnm{Ethan} \sur{Tokar}}\email{etokar@u.rochester.edu}
\equalcont{These authors contributed equally to this work.}

\author[2]{\fnm{Esteban} \sur{Wright}}\email{ewrig15@ur.rochester.edu}
\equalcont{These authors contributed equally to this work.}

\author[3]{\fnm{Paul} \sur{S\'anchez}}\email{diego.sanchez-lana@colorado.edu}
\equalcont{These authors contributed equally to this work.}

\author[4]{\fnm{Rachel} \sur{Glade}}\email{rglade@ur.rochester.edu}
\equalcont{These authors contributed equally to this work.}

\author[5]{\fnm{Alice} \sur{Quillen}}\email{aquillen@pas.rochester.edu}
\equalcont{These authors contributed equally to this work.}

\author*[1]{\fnm{Hesam} \sur{Askari}}\email{askari@rochester.edu}
\equalcont{These authors contributed equally to this work.}

\affil*[1]{\orgdiv{Department of Mechanical Engineering}, \orgname{University of Rochester}, \city{Rochester}, \state{NY}, \postcode{14627}, \country{USA}}

\affil[2]{\orgdiv{Institute for Physical Science and Technology}, \orgname{University of Maryland}, \city{College Park}, \state{MD}, \postcode{20742}, \country{USA}}

\affil[3]{\orgdiv{Colorado Center for Astrodynamics}, \orgname{University of Colorado Boulder}, \city{Boulder}, \state{CO}, \postcode{80303}, \country{USA}}

\affil[4]{\orgdiv{Department of Earth and Environmental Sciences}, \orgname{University of Rochester}, \city{Rochester}, \state{NY}, \postcode{14627}, \country{USA}}

\affil[5]{\orgdiv{Department of Physics and Astronomy}, \orgname{University of Rochester}, \city{Rochester}, \state{NY}, \postcode{14627}, \country{USA}}


\abstract{The interactions of solid objects with granular media is countered by a resistance force that stems from frictional forces between the grains and the media's resistance to inertia imposed by the intruder.  Earlier theories of granular intrusion have suggested an additive contribution of these two families of forces and had tremendous success in predicting resistive forces on arbitrary shaped objects. However, it remains unclear how these forces are influenced by gravitational conditions. We examine the role of gravity on surface impact behavior into cohesionless granular media using hundreds of soft-sphere discrete element simulations, we demonstrate that the outcome of impacts remain qualitatively similar under varying gravitational conditions if initial velocities are scaled with the Froude number, suggesting an underlying law. Using theoretical arguments, we provide reasoning for the observed universality and show that there is a hidden dependency in resistive forces into granular media on Froude number. Following the theoretical framework, we show that Froude number scaling precisely collapses impact trajectories across gravitational conditions, setting the foundation for explorations in granular behavior beyond Earth.}




\maketitle

\section{Main}\label{sec1}

Granular materials are ubiquitous on Earth, taking the form of soil, sand, gravel, pebbles, and boulders. They are also prevalent beyond our planet, present on the surfaces of moons, asteroids, and other planets, exhibiting remarkable diversity in size, shape, and dispersity \citep{walsh2022near, sugita19, walsh19, dellagiustina2019properties, Rozitis_2020, Michikami_2021}. The interest in exploration and exploitation of non-terrestrial bodies, often strewn with this granular media, has considerably increased during the last decade for scientific investigations and economic reasons \citep{galache2017asime}. However, any process that involves interaction with granular media will need to overcome the challenges of understanding its wide range of complex mechanical behaviors. This includes history-dependent strength \citep{schofield1968critical}, flow anisotropy and non-associated flow rules \citep{weinhart2013coarse, sulem1995bifurcation,nedderman1992statics}, and nonlinear and non-local behavior \citep{da2005rheophysics,jop2006constitutive,henann2013predictive}. The extent to which a granular media exhibits these characteristics is intimately dependent upon its physical properties as well as the conditions of the surrounding environment. Morphological and environmental conditions such as packing fraction \citep{vo2020additive}, porosity, and cohesional forces \citep{scheeres2010scaling, mandal2020insights} influence the rheology and flow properties of grains. Meanwhile a particularly important factor in the response of grains on non-terrestrial surfaces is the fact that they experience a different gravitational force compared to grains Earth. This is expected to strongly influence the behavior and rheology of granular media since their strength is proportional to the confining pressure which in turn scales with gravity. Understanding the role of gravity in granular behavior is key with creating a clearer picture of granular mechanics beyond Earth.

The interaction between solid objects and granular media is a common feature of both natural and human-made processes. This general interaction is referred to as the `intrusion problem.'  Excavation \cite{dorgan2015biomechanics}, animal locomotion \cite{hosoi2015beneath}, vehicular locomotion \cite{thoesen2020revisiting}, landing apparatus \cite{ulamec2014landing}, the study of craters \cite{neukum1994crater}, and probing missions on non-terrestrial surfaces \cite{boundarysciadv.abm6229} are only a few examples. Granular media is complex, though, in that it can exhibit a wide range of behavior depending on its properties and local environment. Packed granular media may behave like a solid body, while agitated granular media may demonstrate fluid-like or even gaseous behavior \citep{Jaeger1996}. All of these states may emerge during a surface intrusion scenario. Of particular interest in this work are the low-velocity oblique impacts, resulting from both natural and human-made causes. Impacts on asteroids and other low-gravity surfaces can create gravitational-bound ejecta and debris that re-impact the surface at lower velocities. The media's inertial resistance becomes prevalent in resisting surface intrusion during impacts. It remains an open question how the quasi-static strength and inertial resistance of the media evolve  when gravitational condition changes. Analyses of grain-grain level data on particles interacting with the intruder can reveal how granular arrangement and its variations resist in the intrusion process \cite{miklavvcivc2022sub}. However, a more accessible approach is the study of the impactor's response and linking its behavior to variations in the environmental conditions. Therefore, we examine how granular media resists the intrusion process and how the trajectory of the intruder evolves by creating normalized behavior maps to parse out the role of gravity.

\section{Parameterization of gravity}
Experimental evaluations of intrusion problems under different gravitational conditions are challenging and costly, requiring on-board parabolic flights or drop-towers to override Earth's gravity. However, these experiments are hindered by factors such as variations in apparent gravity during parabolic flights and size limitations for drop tower experiments. Alternatively, computational methods and theoretical approaches can provide significant insights into the dynamics of granular media in other gravities. For this purpose, we employ a Discrete Element Method (DEM) environment in which the granular medium is idealized as a collection of round cohesionless particles bound by gravity and Hertizan contact forces (See methods for details). To hone in on the effect of gravity, we adopt this framework which allows for changing gravity as an independent parameter. The size, shape, coefficient of friction, material properties, and densities of the impactor and grains are key factors in the surface intrusion problem. In addition, the velocity and launch angle of the projectile will affect the granular response. To isolate the effects of gravity, we keep all of these properties constant except for the initial speed $v$ of the projectile which is scaled with gravity based on Froude number, $\text{Fr} = v/\sqrt{g \; r}$ to preserve the ratio of inertial forces to gravitational forces. All other parameters and morphological properties of the granular media remain unaffected. While, Froude number has historical applications in fluid dynamics \cite{Ogilvie1967, Tahara1992, Tahara1996, He2019} and gait analysis \cite{Alexander1976}, it applies here because it reports the ratio of inertial to frictional (and therefore gravitational) forces in a system. In parallel to DEM simulations, we use a theoretical model known as the Dynamic Resistive Force Theory (DRFT) \citep{agarwal2019modeling} to explain our observations and ultimately, present a unified scaling law for low velocity impacts into cohesionless granular media in myriad gravitational conditions.

\section{Impact behavior across gravities}\label{sec2}

\begin{figure*}[h]
    \centering
    \begin{subfigure}[b]{0.54\linewidth}
        \includegraphics[width=\linewidth]{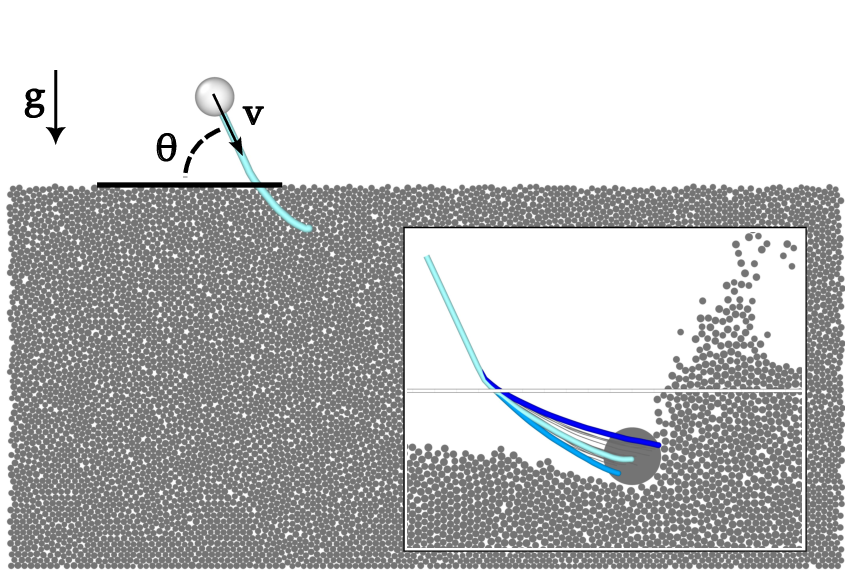}
        \subcaption{ }
        \label{fig:DEMsetup}
    \end{subfigure}
    \begin{subfigure}[b]{0.44\linewidth}
        \includegraphics[width=\linewidth]{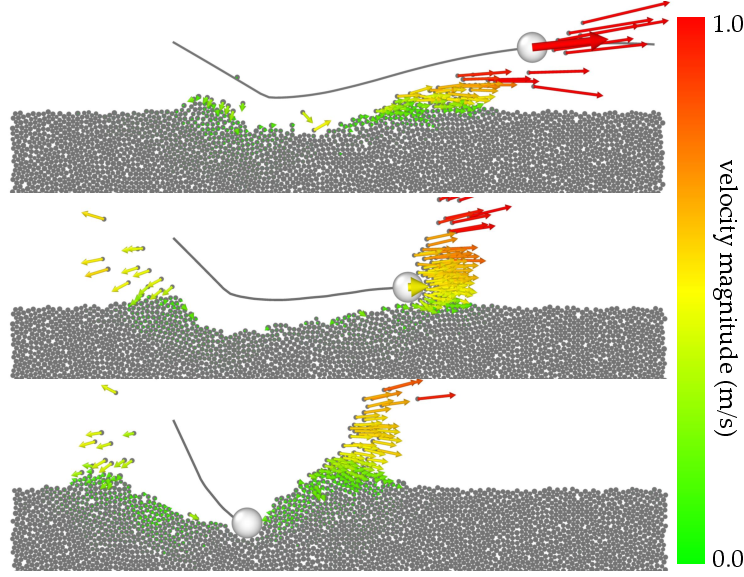}
        \subcaption{ }
        \label{fig:DEMexamples}
    \end{subfigure}
    \caption{(\ref{fig:DEMsetup}) shows conventions for launch velocity and angle. The inset shows variability in path of the projectile when launched from different positions. (\ref{fig:DEMexamples}) shows impact outcomes of ricochets (top), roll-outs (middle), and full-stops (bottom) with colored vectors showing the velocity of the grains and projectile.}
    \label{fig:mainFig1}
\end{figure*}

The DEM environment models an array of surface impacts by an impactor of diameter $D$ in Earth gravity ($g$= 9.81 m/s$^2$), Moon gravity (1.62 m/s$^2$), and asteroid Bennu-like gravity ($6.27\times10^{-5}\;\text{m/s}^2$). The baseline DEM setup is shown in Fig.~\ref{fig:DEMsetup}. Focusing only on the projectile's response to a surface impact as a holistic representation of the granular response, three cases may result as initially defined by Wright et al. \cite{Wright2020, Wright2021}. The impactor may ricochet and bounce from the impact site, roll out of the initial impact crater whilst remaining in contact with the surface, or fully stop in the crater as shown in Fig.~\ref{fig:DEMexamples}. A velocity range of 1 m/s to 7 m/s for Earth's gravity translates to a Froude number range of 4 to 25 which is kept constant for all gravitational environments. We vary the launch angle between 20\degree~and 70\degree~resulting in an array of 143 simulation cases for each gravitational condition. 

% The variations in grain arrangement that develop naturally during settling of grains can result in slight variations in local packing fraction \cite{miklavvcivc2022sub}. This can slightly affect the trajectory and the impact outcome as shown in the inset of Fig.~\ref{fig:DEMsetup}. We repeat each set by moving the initial placement of the impactor by 4 grain diameters to account for these variations and to obtain smoother behavior maps. 

The variations in grain arrangement that develop naturally during settling of grains can result in variations in local packing fraction \cite{miklavvcivc2022sub}. In the case of an impact, this condition can significantly influence projectile trajectory, depending on the projectile-to-grain size ratio of the system \cite{Wright2021}. In this DEM work, the grains are coarse relative to the impactor with a ratio of about 6.7, meaning we expect trajectory influence due to grain configuration. We demonstrate this variation in the inset of Fig.~\ref{fig:DEMsetup}. To account for these variations in the impact studies and to obtain smoother data for the behavior maps, we perform two sets of impact models in each gravity which differ only in the initial placement of the impactor. We use the same arrangement of the settled bed in the repeated set to minimize the additional variations due to local packing. We also use the same settled granular bed obtained under Earth gravity, but depressurized to use for other gravities to maintain packing fraction between cases.

\begin{figure*}[h]
    \centering
    \begin{subfigure}[b]{0.32\linewidth}
        \includegraphics[width=\linewidth,trim={0cm 0 0.5cm 0},clip]{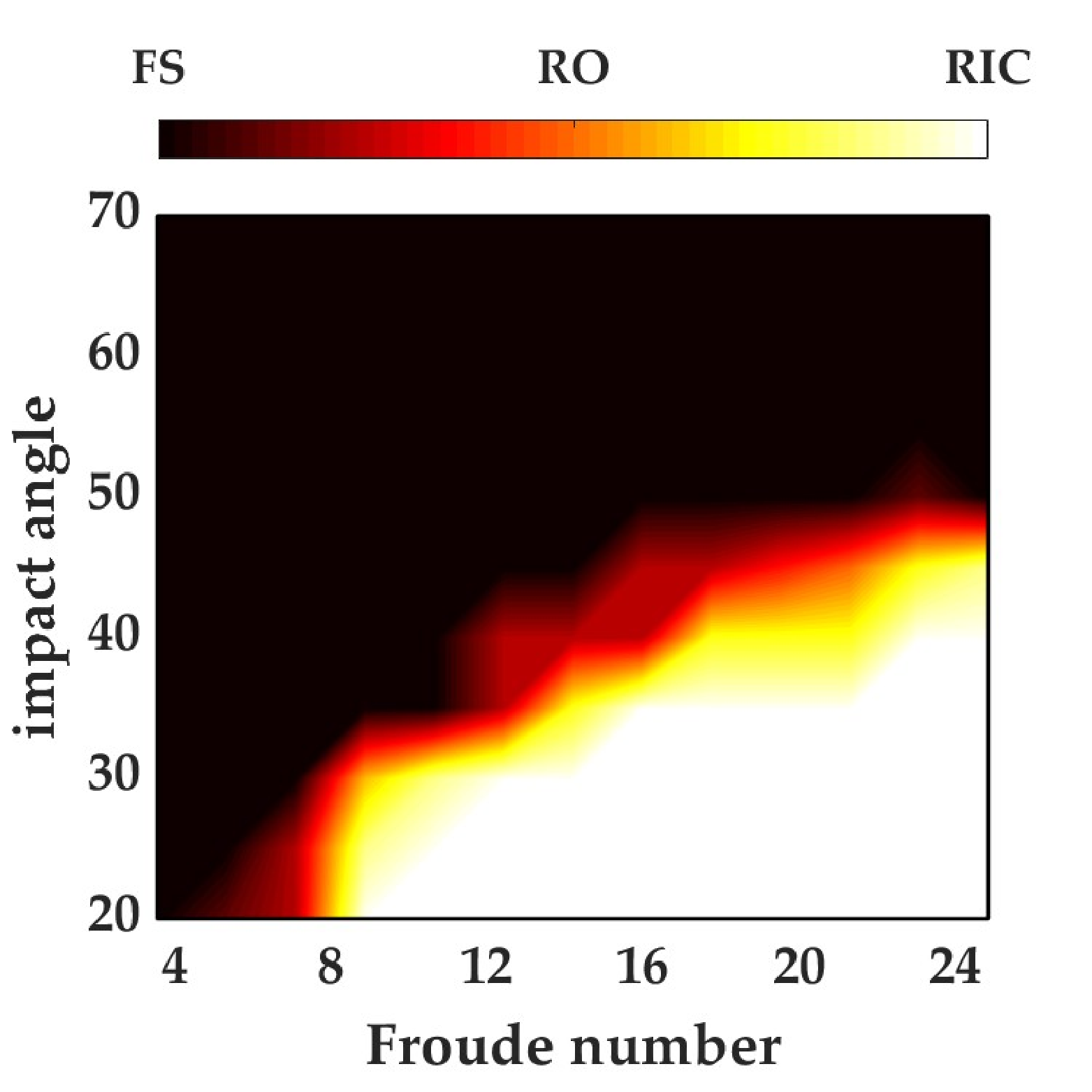}
        \subcaption[]{Earth (9.81 m/s$^2$)}
        \label{fig:EarthMap}
    \end{subfigure}
    \begin{subfigure}[b]{0.32\linewidth}
        \includegraphics[width=\linewidth,trim={0cm 0 0.5cm 0},clip]{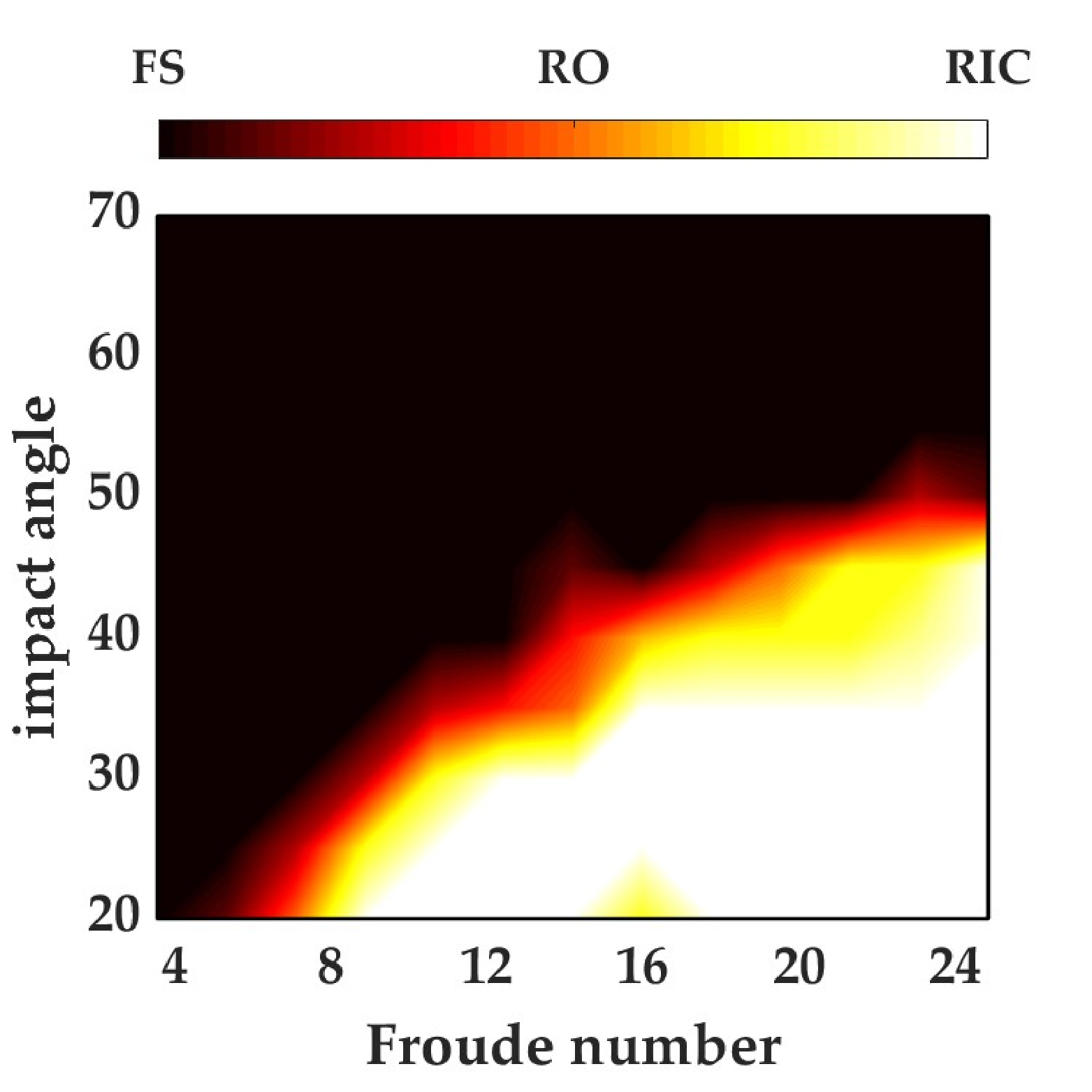}
        \subcaption[]{Moon (1.62 m/s$^2$)}
        \label{fig:MoonMap}
    \end{subfigure}
    \begin{subfigure}[b]{0.32\linewidth}
        \includegraphics[width=\linewidth,trim={0cm 0 0.5cm 0},clip]{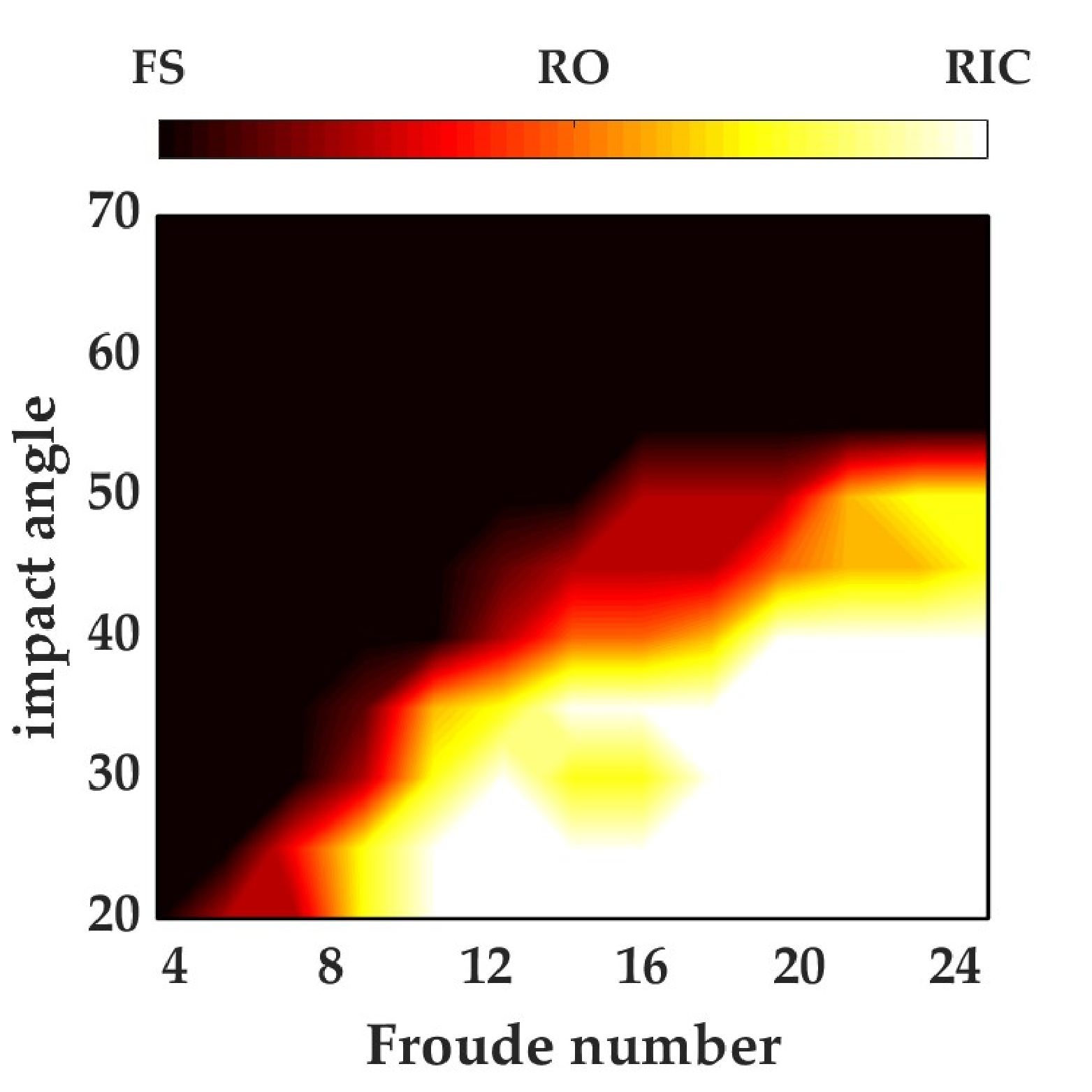}
        \subcaption[]{Bennu ($6.27\times10^{-5}\;\text{m/s}^2$)}
        \label{fig:BennuMap}
    \end{subfigure}
    \caption{Behavior maps produced with DEM simulations for three gravitational conditions. Each map is the averaged from four sets of classifications.}
    \label{fig:AllMaps}
\end{figure*}

We observe remarkable agreement in numerical behavior maps across gravitational conditions as shown in Figs. \ref{fig:EarthMap}-\ref{fig:BennuMap}, which collectively present the outcome of 858 impact simulations. We note that the horizontal intercept of the roll-out boundary region occurs between Froude numbers 5 and 10 and the vertical intercept occurs at or just below 50\degree \: for all behavior maps. These maps corroborate experimental observations by Wright et al. \cite{Wright_2020b} that the ricochet behavior concentrates towards the lower right side of the plot and full-stops in the upper left with the roll-outs observed as a transitional zone in-between. We find similar percentage of roll-out to ricochet region for Earth, Moon, and Bennu gravities as 31.5\%, 32.9\%, and 36.4\%, respectively. There is a trend in these measurements with gravity, but this can be attributed to the marginal decrease in packing fraction resulting from the adjustment of Earth's grain arrangement to accommodate other gravitational conditions.

%We note that the horizontal intercept of the roll-out boundary region occurs between Froude numbers 5 and 10 and the vertical intercept occurs at or just below 50\degree \: for all behavior maps. 

The extensive range of gravitational conditions covered in the DEM results highlights the effectiveness of Froude number scaling in  unifying the impactor response. This raises the intriguing question of why such a high level of universality exists, considering the inherent complexities of granular media and orders of magnitude change in gravity. We propose an explanation by using a reduced-order theory for intrusion into granular media. Furthermore, this theory offers a robust approach to investigating the intrusion problem from a mean-field perspective, effectively mitigating the influence of the discrete nature of the media that is present in DEM simulations. By eliminating variations stemming from this source of uncertainty, a clearer understanding of the impactor's response is obtained.

\section{A gravity-normalized theory for intrusion}\label{subsubsec2}

Inspired by earlier studies on micro-swimming organisms in viscous fluids \cite{gray1955propulsion}, granular Resistive Force Theory (RFT) was proposed to provide a set of rules to calculate the forces arising from the interaction between solid objects and granular media \cite{li2013terradynamics}. In RFT-type models, the granular media is considered a resistant medium that generates forces opposing the motion of intruding objects. This theory, although empirically derived, has demonstrated remarkable effectiveness in predicting resistive forces arising in interactions with granular media in non-inertial regimes \cite{slonaker2017general, luck2017lab, thoesen2019screw}. Its mechanical foundation has been elucidated by establishing connections between theory and measurable physical properties of grains both quasi-static, non-inertial regimes \citep{askari16}. To extend its applicability beyond the quasi-static regime, Dynamic Resistive Force Theory was introduced \cite{agarwal2019modeling}, which is expressed in two dimensions as follows.
\begin{equation}
    \tau_{x,z} = \alpha_{x,z}(\beta, \gamma) \; \text{H}(z) \; z + \lambda \; \rho \; v_n^2 \; n_{x,z}
    \label{eqn:DRFT}
\end{equation}

\begin{figure}[h]
\centering
   \includegraphics[width=0.4\linewidth]{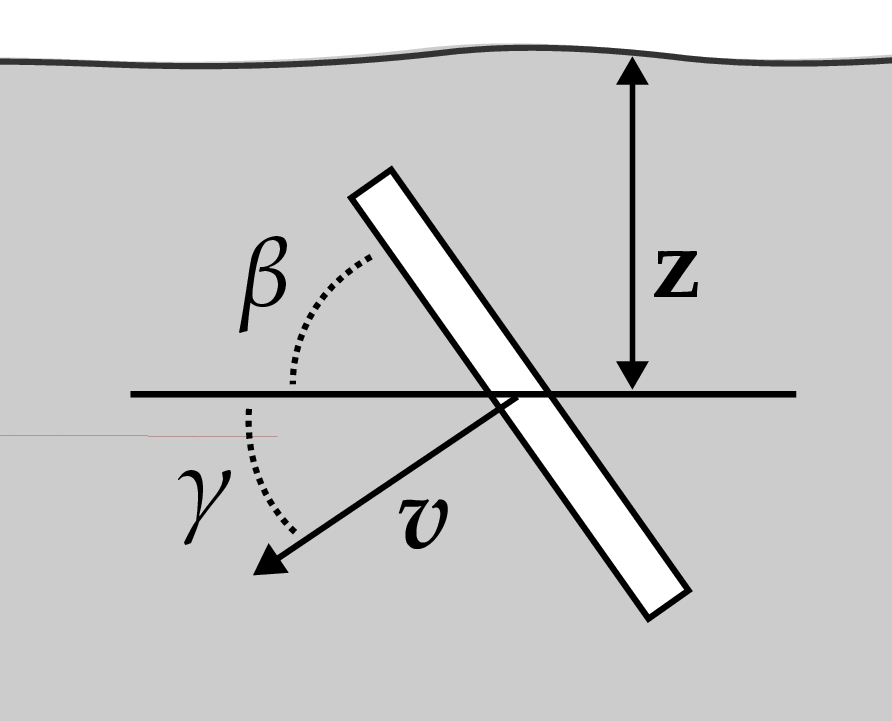}
   \caption{Conventions for orientation angle $\beta$ and velocity direction angle $\gamma$ of a plate intruding at depth $z$ through a resisitive media. These quantities provide the basis for the quasi-static formulation of RFT.}
   \label{fig:DRFT_definitions}
\end{figure}
Consider a 2D space filled with grains defined by the horizontal direction $x$ and vertical direction $z$, the latter aligned with gravity and pointing downward. When an object of arbitrary geometry `intrudes' into this space, any segment of its perimeter moving into the granular media will experience a resistive traction $\tau_{x,z}$ where subscripts denote the  traction components. Equation \ref{eqn:DRFT} includes both the quasi-static and inertial components of the resistance to motion. The quasi-static resistance arises from friction between the grains and depends on the orientation angle $\beta$, velocity angle $\gamma$, and depth $z$ (pictured in Fig. \ref{fig:DRFT_definitions}). This term is determined by empirical resistive force data that form a vector-valued quantity $\alpha_{x,z}(\beta, \gamma)$ with units of traction per depth (see Supplements for details). A Heaviside function $H(z)$ ensures segments above the surface do not experience a force. The inertial resistance is calculated using a fitting parameter $\lambda$, bulk density of the granular media $\rho$, and has a quadratic dependence on the speed $v_n$ in the normal direction of the segment defined by $n$. By integrating over all infinitesimal areas while excluding areas moving away from the media, the resistive force on the entire object can be obtained.

While the weight's dependence on acceleration of gravity is straightforward, additional considerations are required to use Equations \ref{eqn:DRFT} in different gravitational conditions. Frictional forces are influenced by pressure, which is influenced by both gravity and depth \cite{Roth2021Intrusion, Agarwal2023Mechanistic}. While Equation \ref{eqn:DRFT} explicitly includes the depth variable $z$, the dependence on gravity is implicitly incorporated into $\alpha_{x,z}$. Through theoretical and continuum understanding of RFT, we recognize that $\alpha_{x,z}$ originates from measurable quantities of the granular media and can be reproduced if mass density and coefficient of friction of the grains are known \cite{askari16}. Thus, to remove gravity dependence from $\alpha_{x,z}$, we can envision an Earth gravity-normalized $\alpha^{0}_{x,z}$ defined as $\alpha_{x,z}/g$. 

%We also know that the strength in cohesionless granular media increases with pressure and pressure is primarily due to gravity \cite{Agarwal2023Mechanistic}.
%Since frictional forces scale with gravity, the dependence of  $\alpha_{x,z}$ on gravity can be removed by envisioning an 
%Therefore, it can be deduced that $\alpha_{x,z}$ depends on weight density of the media and a gravity-normalized $\alpha^{0}_{x,z}=\alpha_{x,z}/g$ can be envisioned.

% The inertial term, however, is not dependent on gravity since density and velocity are unaffected by it and $\lambda$ may be assumed to be gravity-independent.

The inertial term is not dependent on gravity 
%since velocity is an independent parameter 
and $\lambda$ may be treated as gravity-independent. In addition, preserving packing fraction keeps bulk density constant as gravity changes. In order to make the second term gravity explicit, we rewrite it in terms of Froude number by  $v_n^2 = g r \text{Fr}^2$. Combining our new definitions for quasi-static and inertial terms, and dividing the results by Earth gravity, we find a gravity-normalized traction written as
\begin{equation}
    \tau^0_{x,z} =  \frac{\tau_{x,z}}{g} = \alpha^0_{x,z}(\beta, \gamma)  \text{H}(z) |z| + \lambda  \rho  r \text{Fr}^2 \; n_{x,z}.
    \label{eqn:DRFT_mod}
\end{equation}
This equation shows that, the normalized tractions remain constant in an arbitrary gravitational condition $g^*$ if the Froude number matches between these conditions. Equation \ref{eqn:DRFT_mod} coupled with an initial Froude number, an expression for the area in contact with the granular media, and the mass of the impactor are sufficient to form the Equations of Motion (EOM) for the impactor. We can then solve the EOM in an arbitrary gravitational condition to produces an independent theoretical analysis.
%based on DRFT analysis which can then be compared to DEM results. 

% This is where we present the vertical drop results. 

%In order to gain a better understanding of the more complex problem of explaining the universal impact response, we initially tackle a simpler scenario of 
\section{Universal behavior and vertical impacts}
We can consider a simpler case of intrusion in which the intruder vertically approaches the surface similar to common procedures  for development of vertical force laws in Earth gravity \cite{tsimring05, gpoldman08, katsuragi13}. The purpose of this exercise is to demonstrate how Equation \ref{eqn:DRFT_mod} can be used to study impact trajectories in different gravities.  Placing a flat plate initially at the surface, and imposing an initial Froude number and driving it downwards into the media, we can write the EOM for Earth and non-Earth accelerations of gravity ($g$ and $g^*$, respectively) as
   \begin{align}
    (\alpha_z^0 g z + \lambda \rho \Dot{z}^2)&A  -m g = m \ddot{z}  \label{eqn:EOMg}\\
    (\alpha_z^0 g^* z^* + \lambda \rho \Dot{z}^{*2})&A  -m g^* = m \ddot{z}^* 
    \label{eqn:EOMg*}
    \end{align}
where $z$ and $z^*$ represent the position the plate as a function of time $t$ in Earth and non-Earth conditions and dot notation denotes time derivatives. While the expressions $z$ and $z^*$ cannot be obtained exactly due to the nonlinearities in the EOM, we can consider an unknown function $z$ which is the solution to Equation \ref{eqn:EOMg}. Using a change of variable for time as $t=t^* \sqrt{g^*/g}$, Equation \ref{eqn:EOMg*} takes the same form as Equation \ref{eqn:EOMg} with $t^*$ being the new measure of time (see Supplementary Materials for details). With the initial velocity scaled as $v_0^*=v_0\sqrt{g^*/g}$ so that the initial Froude number is the same, we can show that $z(t)=z^*(t^*)$. This  means both plates will pass the same points in vertical space but the plate in lower gravity will require more time to cover that same trajectory. This finding from DRFT is consistent with similar conclusions found using other phenomenological models for granular behavior \citep{Roth2021Intrusion,Sunday2022a, Sunday2022b}.

% The solution to EOM provides the trajectory and rotation (*** we haven't gone deep into rotation EOM ***) of the impactor before, during, and after the impact. Since resistive forces are presented in traction form, the contact area is needed  to write the EOM. Contact area itself is a function of depth which can make a complicated set of equations to solve exactly. Considering a vertical impact, the solution is straightforward. Consider a round impactor with diameter $D$ moving into the surface with a velocity $v$. Assuming the lowermost part of the impactor positioned at depth $z$, the net resistive force will have a form ....? 

\section{Universal behavior and oblique impacts}
The EOMs become increasingly complex for the full problem of a round impactor in oblique impact since the area in contact depends on depth and angles $\beta$ and $\gamma$  evolve due to the rotation of the projectile. However, they can be solved numerically once inputs for the DRFT Equation~\ref{eqn:DRFT_mod} are set to closely match the discrete model as detailed in the supplementary materials. We generate trajectories and behavior maps based on the theoretical equations and use the same bounds for Froude number and launch angle but at smaller increments. 
%In contrast to the 143 simulations required per behavior map data set from the discrete model, the increased efficiency of the DRFT model and smaller increment affords 3575 different impact scenarios to be studied. 

\begin{figure*}[b]
    \centering
    % \begin{subfigure}[b]{0.6\linewidth}
    %     \includegraphics[height=1.3in, trim={1.2cm 0cm 2cm 1.5cm}, clip]{drft_imp_main.png}
    %     \subcaption[]{}
    %     \label{fig:DRFT_setup}
    % \end{subfigure}
    \begin{subfigure}[b]{0.54\linewidth}
    \includegraphics[width=\linewidth, trim={3cm 0cm 0cm 0cm}, clip]{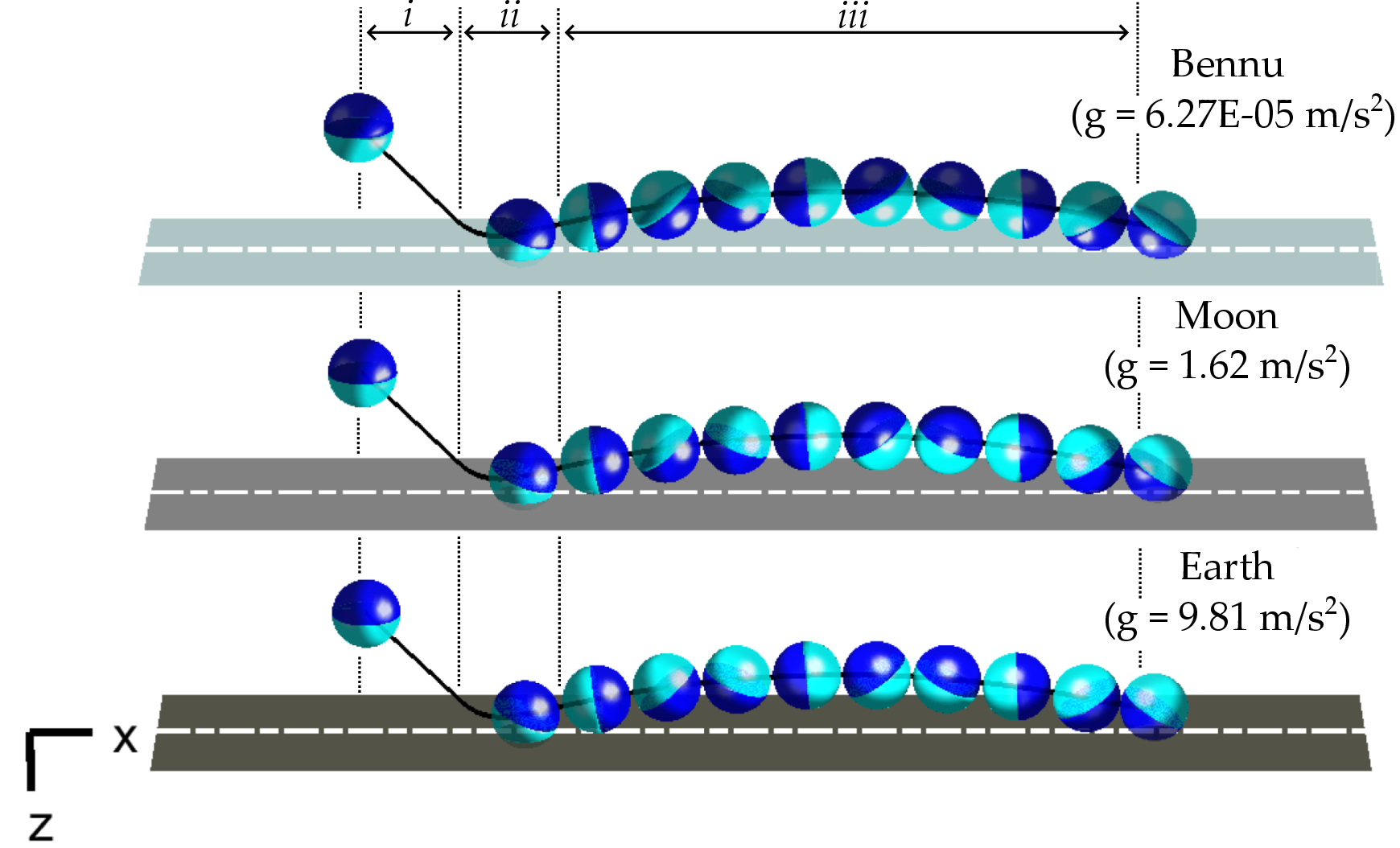}
        \subcaption[]{}
        \label{fig:DRFTexample}
    \end{subfigure}
    \begin{subfigure}[b]{0.4\linewidth}
        \includegraphics[width=\linewidth, trim={0 0 0 0}, clip]{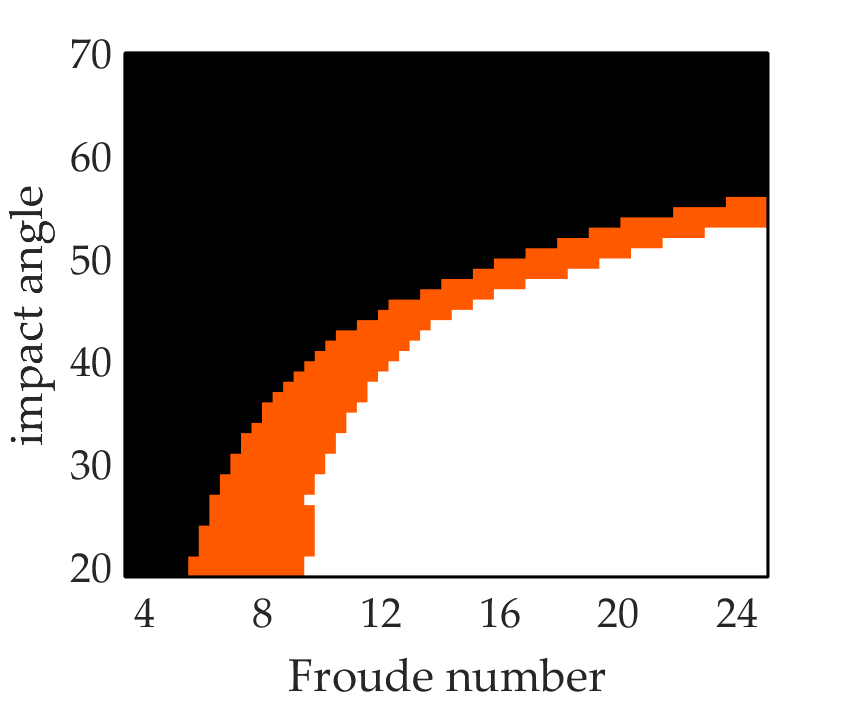}
        \subcaption[]{}
        \label{fig:DRFTcompare}
    \end{subfigure}
    % \begin{subfigure}[b]{0.37\linewidth}
    %     \includegraphics[width=\linewidth, trim={0 0 0 0}, clip]{theory_v2.eps}
    %     \subcaption[]{}
    %     \label{fig:DRFT_stageii}
    % \end{subfigure}
    \caption{ (\ref{fig:DRFTexample}) shows ricochet events modeled at the same Froude number and launch angle in Bennu, Moon, and Earth conditions. (\ref{fig:DRFTcompare}) presents a behavior map constructed with the DRFT model. Black represents full-stop, orange represents roll-outs, and white represents ricochets.}
    \label{fig:AllDRFT}
\end{figure*}

The theoretical solution reveals that the trajectory for a pair of Froude number and launch angle is identical between gravities in oblique impact. This is showcased in Fig.~\ref{fig:DRFTexample} for a 45\degree~launch at a Froude number of 24.87 which is expected to produce a ricochet based on our DEM results in Fig.~\ref{fig:AllMaps}. This equates to 7.00 m/s initial speed for Earth and 2.84 m/s, and 1.77 cm/s for  Moon, and Bennu-like gravities, respectively. 
%These impact conditions are a deliberate choice because the extended airborne duration after interaction with the granular media will allow any variations between the simulations to amplify and be more observable. As a result, the behavior maps produced in each gravity are identical to the one shown in Fig.~\ref{fig:DRFTcompare}. 
The emergence of the same trajectory results in a perfect collapse of theoretical behavior maps shown in Fig.~\ref{fig:DRFTcompare} which is in agreement with DEM maps of Fig.~\ref{fig:AllMaps}. This agreement is achieved by the force law of Equation \ref{eqn:DRFT_mod} and rectifies the need to use different pre-maximum and post-maximum force law as done in \citep{Wright2020}.

%The success of the empirical DRFT model at producing behaviors similar to DEM is also significant as it represents a single empirical framework capturing both the pre-maximum intrusion and post-maximum intrusion phases of an impact. This was a challenge noted by \cite{Wright2020} that was only resolved using different empirical models for each phase.

%to the Froude number scaling, and one of these maps is shown in Figure \ref{fig:DRFTcompare}. The roll-out region is much clearer in the higher resolution data, but the overall characteristics of the map compared to those of Figure \ref{fig:AllMaps} are exceedingly similar. 

%and we first solve a 45\degree~impact at a Froude number of 24.87 which is expected to produce a ricochet based on our DEM models. This Fr equates to 7.00 m/s initial speed for Earth and 2.84 m/s, and 1.77 cm/s for  Moon, and Bennu-like gravities. These impact conditions are a deliberate choice because the extended airborne duration after interaction with the granular media will allow any variations between the simulations to amplify and be more observable. Indeed, the impact solved with DRFT produces a ricochet in each gravitational environment and, as shown in Figure \ref{fig:DRFTexample}, the impacts produce precisely the same trajectory, just over different durations of time. 

%To obtain more insights into the inner-workings of such a remarkable collapse of trajectories and why it should be expected, we take a closer look into the EOM. 

There are three distinguishable stages in the trajectories shown in Fig.~\ref{fig:DRFTexample}; (\textit{i}) the projectile motion prior to contacting the resistive media, (\textit{ii}) the period of interaction with the media, and (\textit{iii}) the post-impact projectile motion . Considering  a round impactor launched with initial velocity $v_0$ and a launch angle of $\theta$ in a two dimensional space defined by $x$ and $z$, with $z$ pointing down in the direction of gravity, the EOM for the center of mass in Earth gravity is $x= v_0 \cos(\theta) t$ and $z=v_0 \sin(\theta) t+ 1/2 g t^2$ in the xz plane.  If acceleration of gravity changes to $g^*$ but the launch angle $\theta$ is kept the same, a new initial velocity  $v^*_0=v_0\sqrt{g^*/g}$ should be considered to keep the initial Froude number constant. In the other gravity $g^*$, the EOM becomes $x^*= v_0^* \cos(\theta) t$ and $z^*=v_0^* \sin(\theta) t + 1/2 g^* t^2 $. If we use a linear scaling of time as $t^*= \sqrt{g/g^*} t$  similar to the vertical intrusion case in the EOM for $g^*$ environment, we find 
%    \begin{align}
%    x^*(t^*) &= v_0  \cos(\theta)  \left(\sqrt{\frac{g^*}{g}} (\sqrt{\frac{g}{g^*}} t) \right)  \text{and} \label{eqn:EOM1}\\
%    z^*(t^*) &= v_0  \sin(\theta)  \left(\sqrt{\frac{g^*}{g}} (\sqrt{\frac{g}{g^*}} t) \right) + \frac{1}{2} g^* (\sqrt{\frac{g}{g^*}} t) ^2 .
%    \label{eqn:EOM2}
%    \end{align}
%With some simple steps, Equations \ref{eqn:EOM1} and \ref{eqn:EOM2} 
that the EOM in $g^*$ condition reduce to Earth's EOM and then produce the equalities $x^*(t^*)=x(t)$ and $z^*(t^*)=z(t)$. This results in similar findings to the vertical intrusion case where the spatial response coincides but the temporal domain follows a scaling of $t^* = t \sqrt{g/g^*}$. Therefore, the universality of the trajectory is proven for stage (\textit{i}). Note that Froude number is not constant during the stage (\textit{i}) due to gravitational acceleration, but the evolution of Froude number remains identical between gravities.

% \begin{figure}[t]
%     \centering
%     \includegraphics[width=1\linewidth]{theory_v2.eps}
%     \caption{An illustration of two projectiles in different gravities $g$ and $g^*$. The frame at $t_1$/$t^*_1$ is the first point of contact whereas $t_2$/$t^*_2$ occur further into the intrusion. This figure accompanies the theoretical discussion that shows how Froude number evolves identically across gravitation conditions.}
%     \label{fig:DRFT_stageii}
% \end{figure}

%We can absorb the $\sqrt{g^*/g}$ prefactor into time $t$ and introduce a new gravity-scaled time as $t^*$ and then the EOM will be simplified to 
%    \begin{align}
%    x &= v^*_0 \cos(\theta ) t^* \label{eqn:EOM_unifieda}\\
%    z &= v^*_0 \sin(\theta ) t^* + \frac{v^2_0}{2\text{Fr}^2} t^{*2}.
%    \label{eqn:EOM_unifiedb}
%    \end{align}
%This means that the trajectory of the throw before contacting the surface is set by the choice of throw angle $\theta$, initial Fr and changing gravity while keeping Fr a constant will not alter the points in space passed by the impactor to reach the surface. 
%The same argument applies to the trajectory after a ricochet event since the impactor is no longer interacting with the granular media. 

%Stage (\textit{ii}) of the trajectory begins when the first point on the perimeter of the impactor contacts the resistive granular media (Fig.~\ref{fig:DRFT_stageii}). 

\begin{figure}{t}
\centering
   \includegraphics[width=0.6\linewidth]{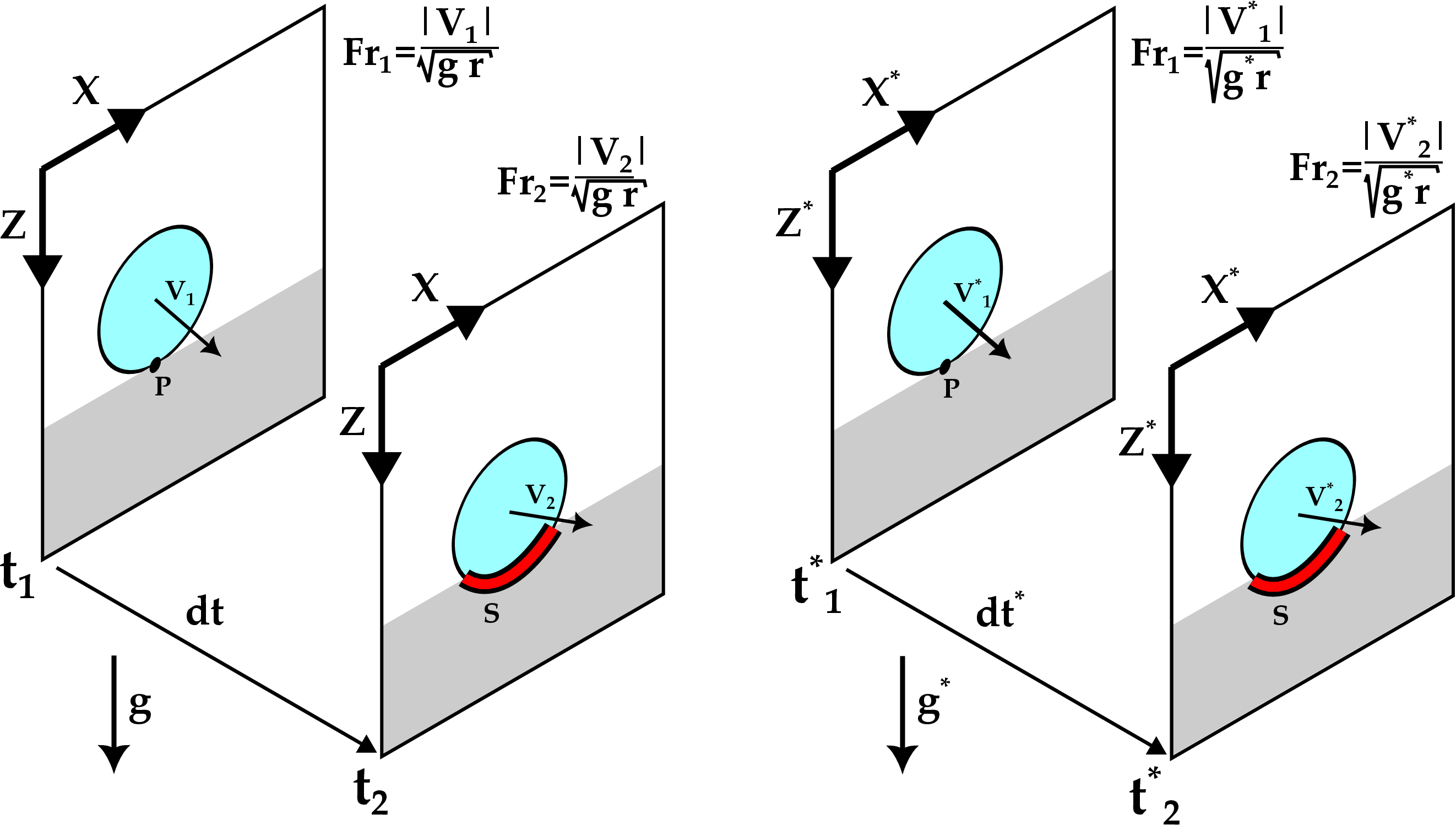}
   \vspace{1em}
   \caption{an illustration of two projectiles in different gravities $g$ and $g^*$. The frame at $t_1$/$t^*_1$ is the first point of contact whereas $t_2$/$t^*_2$ occur further into the intrusion. This figure accompanies the theoretical discussion that shows how Froude number evolves identically across gravitation conditions.}
   \label{fig:DRFT_stageii}
\end{figure}
Based on our arguments for stage (\textit{i}), the same point of the impactor will make the first contact across gravitational conditions to initiate stage (\textit{ii}). This will happen at different times $t$ and $t^*$ as illustrated in Fig.~\ref{fig:DRFT_stageii}. The gravity-normalized traction $\tau^0_{x,z}$ experienced by that point $P$ will also be the same according to Equation \ref{eqn:DRFT_mod}. If we consider a consecutive frame in time different by $dt$ in Earth's gravity and $dt^*$ in non-Earth, the principle of impulse and momentum  between these two frames in Earth's gravity states that 
\begin{equation}
    dt^* \int_{S} \tau^0_{x,z} g^* dA = (Fr_{[t^*+dt^*]} \Vec{u}_2^*-Fr_{[t^*]} \Vec{u}_1 )\sqrt{g^* r} 
    \label{eqn:momentumg*}
\end{equation}
and for Earth gravity as 
\begin{equation}
    dt \int_{S} \tau^0_{x,z} g dA = (Fr_{[t+dt]} \Vec{u}_2-Fr_{[t]} \Vec{u}_1 )\sqrt{g r} 
    \label{eqn:momentumg}
\end{equation}
Considering the same scaling in time as $t^*=t\sqrt{g/g^*}$ and substituting it in Equation \ref{eqn:momentumg*}, we obtain 
\begin{equation}
    dt \int_{S} \tau^0_{x,z} g dA = (Fr_{[t^*+dt^*]} \Vec{u}_2^*-Fr_{[t^*]} \Vec{u}_1 )\sqrt{g r} .
    \label{eqn:momentumg*1}
\end{equation}
We know that $Fr_{[t^*]} \Vec{u}_1 = Fr_{[t]} \Vec{u}_1$ based on the arguments of stage (\textit{i}) and the left hand side of Equations \ref{eqn:momentumg*1} and \ref{eqn:momentumg} are the same. Therefore, we can deduce that $Fr_{[t^*+dt^*]} \Vec{u}_2^*= Fr_{[t+dt]} \Vec{u}_2$. This means that $Fr_{[t^*+dt^*]} = Fr_{[t+dt]}$ and $\Vec{u}_2^*=  \Vec{u}_2$, and furthermore that the surface $S$ must also evolve the same way between the two time frames. Any subsequent time frame is subject to the same argument. As a result, the evolution of Froude number, direction of velocity, and the contact surface area will be the same throughout stage (\textit{ii}), across gravitational conditions but over different time scales. This scaling also applies to the angular velocity of the impactor that develops due to the interaction with the media. Therefore, not only will the trajectories match, but the rotations will also match between gravities. Since the state of motion at the departure of stage (\textit{ii}) will be the same across gravites, it then follows via the arguments for stage (\textit{i}) that trajectories throughout stage (\textit{iii}) must also coincide.

\section{Discussion}\label{sec12}

The collective results from DEM simulations accompanied by the theoretical explanation provide a robust case and firm reasoning for the effectiveness of Froude number scaling in impacts and interactions with cohesionless granular media across gravitational conditions. While DEM results show a near-complete collapse of behavior maps, the theoretical explanation uses a non-discrete fundamental description of granular intrusion which is wholly independent from the DEM approach and free from the variabilities inherent in a discrete model. The theoretical arguments unveil how and why Froude number scaling works, and how it spatially collapses all stages of a projectile's trajectory across gravitational conditions. Furthermore, it is shown that the temporal domains across gravities can be related to each other with a simple scale factor derived from Froude number.

Our findings of unified behavior stem from assumptions that the media is cohesionless and that the compaction level of the media at different gravities is the same as Earth's compaction, the latter assisting in the isolation of gravity as a sole source of influence. Further considerations should be made in the presence of cohesion and various compactions as these conditions may be more representative of the physical conditions on non-terrestrial surfaces \citep{scheeres2010scaling, sanchez2020cohesive}. Nevertheless, our study contributes to the fundamental understanding of gravity's role in granular impacts and paves the way for exploring this phenomenon in three dimensions and in the presence of more complex and representative granular characteristics such as cohesion, complex grain geometries and size distributions, or variable packing fractions.

%Our study advances the understanding of the influence of gravity on surface impacts into cohesionless granular media. Through DEM simulations, we have demonstrated that scaling the impactor velocity with the gravitational acceleration of the environment to maintain a constant Froude number yields consistent impact outcomes. 
%This finding highlights the crucial role of Froude number scaling in unifying the trajectory of the impactor throughout its interaction with the granular media and while it is in free flight. 
%Our study provides a fundamental understanding of the role of gravity in surface interactions with the granular media  that explains the observed consistency in DEM simulations across gravities. We presented a major outcome from the theoretical study that shows Froude number scaling of velocity based on present acceleration of gravity unifies trajectory of the impactor before, during, and after its interaction with the granular media. However, the duration of time needed for the impactor to cover a certain trajectory is dependent on acceleration of gravity of the environment.

\section{Methods}\label{sec11}

\subsection{LAMMPS discrete simulation}

LAMMPS is an open-source molecular dynamics solver with a broad range of applications from large-scale to nano-scale. For granular media, it is a platform for discrete so-called soft-sphere discrete element method simulations where each grain is modeled as a sphere and interactions with its neighbors are solved by applying force proportional to the level of overclosure between them. 

The models in this work use a `hertz/material' definition for solving interactions. For two interacting particles, identified by $i$ and $j$, the normal elastic component of the force ($F_{ne}$) between them is computed in terms of the effective elastic modulus $E_{eff}$, the effective particle radius $R_{eff}$, and the overlap between the two particles $\delta_{ij}$ by the expression

\begin{equation}
    F_{ne} = \frac{4}{3} E_{eff} R_{eff}^{1/2} \delta_{ij}^{3/2} \boldsymbol{n}
\end{equation}

A damping coefficient based on work by Tsuji et al. \cite{tsuji1992lagrangian} is computed as a function of the coefficient of restitution. Finally, for tangential response, a Coulombic friction model is imposed that is based on a user-input friction coefficient. The details of the values selected for these properties as well as the construction of the impact model are discussed in the next section.

\subsection{Numerical implementation of DRFT}

To solve an impact with idealized conditions of DRFT, Eqn. \ref{eqn:DRFT_mod} must be solved. An exact solution is not possible, as the velocity-squared dependence in the inertial component creates a non-linear differential equation. A numerical solution can be achieved, though, by discretizing the circular projectile into a number of linear segments and incrementally solving for its path through the resistive medium. These increments are governed by a fixed time step and for each step the projectile's overall resistive forces, gravitational forces, and inertia are balanced.

The spatial discretization of the projectile's surface and the temporal discretization of the solution invite potential sources of error to the model. Circle discretization was tested between 10 and 100 segments, yet model response was observed to saturate at greater above 20 segments. 20 segments therefore became the model's baseline, though it is notable that changing the number of segments only ever minorly influenced results as the DRFT solution further discretizes segments if they are only partially submerged in the granular surface. Timesteps between 1e-3s and 1e-7s were tested, and the model was observed to saturate above 1e-5s which was then also set as baseline. Overall wall time of the model is extremely brief and with the baseline setup solving for an impact that takes place over 0.1s, the wall time was less than 1s. 

\backmatter

\bmhead{Supplementary information}

The included Supplementary Information is a single PDF file that provides further details needed for the reproducability of the results of this work.

\bmhead{Acknowledgments}

We acknowledge the financial support from NASA grant number 80NSSC21K0143 and the support from the University of Rochester's intramural grant. We are also thankful to Mokin Lee for his contributions to this study.

\section*{Declarations}

The authors have no conflicts of interest to declare.

\section*{Declarations}

The authors have no conflicts of interest to declare.

\section*{Author contributions}
H.A. designed and supervised the project. P.M. and E.T. performed and analyzed the simulations. P.M. performed the theoretical analyses and implementations under guidance of E.W., P.S., R.G., A.Q., and H.A. All authors contributed to the discussions and results of the manuscript. 
\pagebreak

% %% UNCOMMENT THIS BLOCK NEW BBL MUST BE GENERATED
% % \bibliographystyle{sn-nature}
% \bibliography{references.bib}% common bib file
% %% if required, the content of .bbl file can be included here once bbl is generated
% % \input sn-article.bbl

%% PASTE BBL FILE CONTENTS BELOW

\end{document}